\documentclass[prd,reprint,twocolumn,notitlepage,preprintnumbers,superscriptaddress]{revtex4}
\usepackage{wrapfig}
\usepackage{verbatim}
\usepackage{amsmath, multirow, amssymb, wasysym, gensymb}
\usepackage{graphicx}
\usepackage{amsfonts}
\usepackage{graphicx}
\usepackage{graphics}
\usepackage[ngerman, english]{babel}
\usepackage{float}
\usepackage{youngtab}
\usepackage{subfigure}
\usepackage[utf8]{inputenc}
\usepackage[usenames,dvipsnames]{xcolor}
\usepackage[normalem]{ulem}
\usepackage{color}

\begin{document}
\preprint{DO-TH 15/16}
\title{A common origin of $R_K$ and neutrino masses}
\author{H. P\"as}
\affiliation{Fakult\"at f\"ur Physik, Technische Universit\"at Dortmund, 44221 Dortmund,
Germany}
\author{E. Schumacher}
\affiliation{Fakult\"at f\"ur Physik, Technische Universit\"at Dortmund, 44221 Dortmund,
Germany}

\begin{abstract} \noindent
The same leptoquarks that explain the recently observed anomaly in $R_K$ can generate naturally small Majorana neutrino masses at one-loop level through mixing with the standard model Higgs boson. This is particularly relevant in models with at least two leptoquarks contributing to $b \rightarrow s ll$ transitions.
\end{abstract}

\maketitle

\section{Introduction}
Recently, the LHCb collaboration announced a $2.6\,\sigma$ deviation from the standard model (SM) in the observable $R_K$ \cite{Hiller:2003js}, which measures the ratio of $B\rightarrow K \mu \mu$ over $B \rightarrow K ee$ decays, implying a breaking of lepton universality. 
The reported result amounts to  \cite{Aaij:2014ora}
\begin{align}
R_K^{\text{LHCb}} = 0.745 \pm \substack{0.090 \\ 0.074} \pm 0.036\,,
\end{align}
as opposed to the SM prediction $R_K^{\text{SM}} = 1.0003 \pm 0.0001$.

The measurement of $R_K$ has caused some excitement in the field, including the revival of TeV scale leptoquarks which explain the anomaly with modified $b \rightarrow s ll$ transitions \cite{Hiller:2014yaa,Hiller:2014ula,Varzielas:2015iva,Sahoo:2015qha,Sahoo:2015fla,Alonso:2015sja,deBoer:2015boa,Becirevic:2015asa,Calibbi:2015kma,Gripaios:2014tna} (cf. Fig. \ref{fig:nunu} (a)).
Conveniently, constraints from $B$-physics force the leptoquarks into a testable range, $1\,\text{TeV} \lesssim M \lesssim 50\,\text{TeV}$, with an upper bound on their mass dictated by the $B_s - \overline{B}_s$ mixing phase \cite{Hiller:2014yaa}.

In this letter we explore the possibility of such TeV scale leptoquarks being responsible for low scale neutrino masses, thereby connecting two currently unresolved phenomena of the SM. The idea of leptoquarks as the origin of neutrino masses has been considered before \cite{Hirsch:1996ye,Mahanta:1999xd,AristizabalSierra:2007nf,Babu:2010vp,Kohda:2012sr,Cai:2014kra,Sierra:2014rxa,Helo:2015fba}, but previous attempts expected lighter leptoquarks with small couplings or involve two loops, where one of the leptoquarks cannot be linked to the $B$ anomalies as it couples only to up-type quarks. Explaining $R_K$, on the other hand, requires sizable couplings to down-type quarks and charged leptons, posing a challenge to combine the seemingly distinct observables.

The paper is structured as follows:
In Sec. \ref{secRK} we review how $R_K$ is modified by two scalar leptoquarks, 
identified by their $SU(3)\otimes SU(2)\otimes U(1)_Y$ quantum numbers, respectively, as
$(3,2)_{1/6}$ and $(3,3)_{-1/3}$, simultaneously contributing to $b \rightarrow s ll$. In Sec. \ref{secmasses} we then estimate the effect of these leptoquarks on neutrino masses and explore the possibilities to accommodate the small neutrino mass scale. Some of these scenarios are elaborated in Sec. \ref{secsetup} with a Froggatt-Nielsen (FN) inspired flavor model and a numerical example. We conclude the discussion in Sec. \ref{secconc}.

\section{Explaining $R_K$ with scalar Leptoquarks \label{secRK}}
\begin{figure}[tbh]
\includegraphics[width=0.6\textwidth]{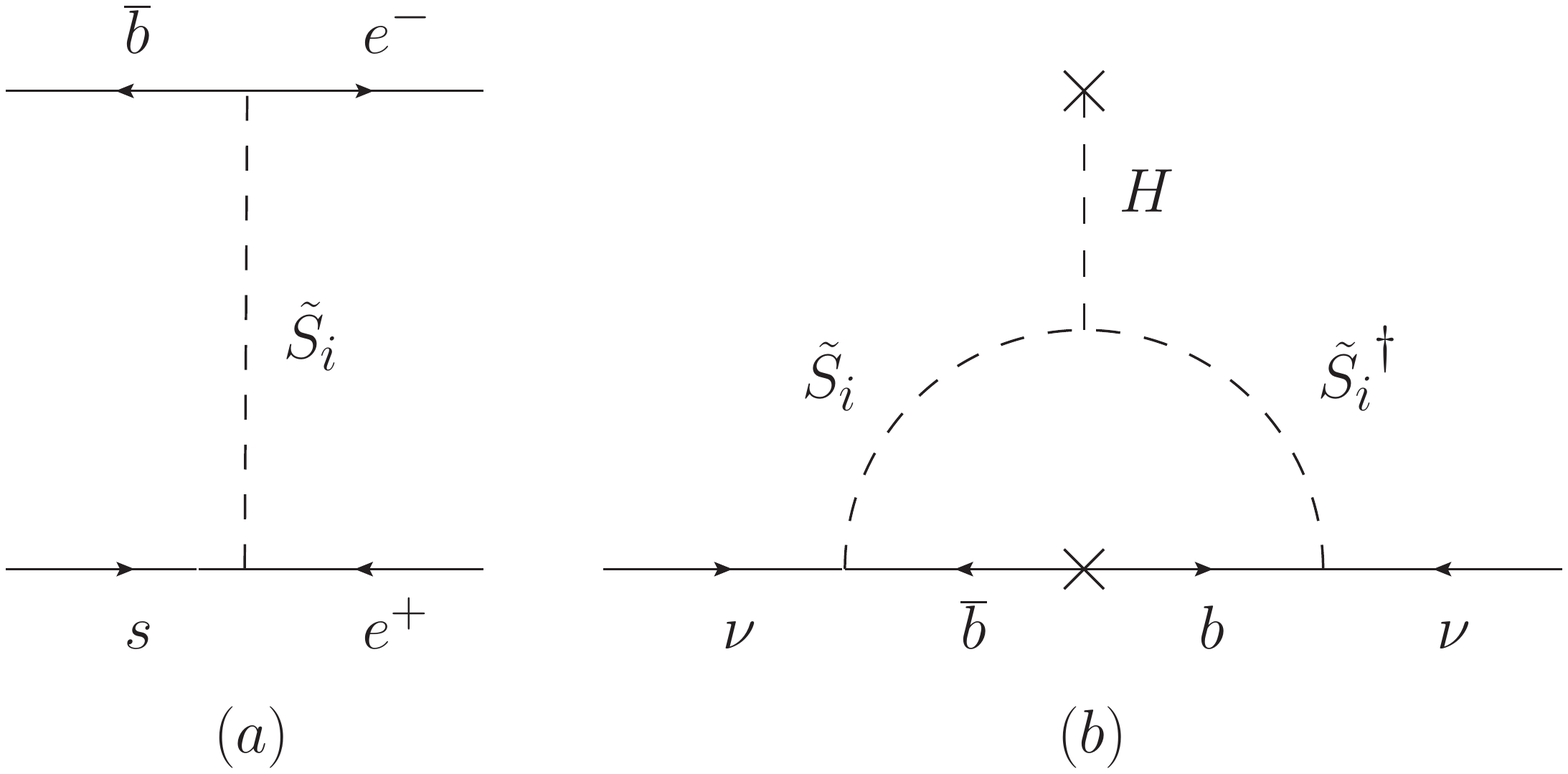}
\vspace{-3.5cm} 
\caption{(a) $\overline{b} \rightarrow \overline{s}l^+l^-$ transition mediated by scalar leptoquarks. $\tilde{S_i}~(i=1,2)$ denotes the leptoquark mass eigenstates defined in Eq. (\ref{massstates}). (b) One-loop level Majorana neutrino mass induced by Higgs-leptoquark mixing. }
\label{fig:nunu}
\end{figure}

As shown in \cite{Hiller:2014yaa}, the deviation in $R_K$ is best explained by (axial) vector operators that, unlike scalar operators, affect $B \rightarrow K ll$ but leave $B_s \rightarrow ll$ unchanged.
Due to Fierz rearrangement 
\cite{Davidson:1993qk}, the desired vector operators are induced by scalar leptoquarks with electric charge $2/3$ or $-4/3$ that couple to down-type quarks and charged leptons. In light of neutrino mass generation we focus only on the $2/3$ leptoquarks $(3,2)_{1/6}$ and $(3,3)_{-1/3}$, whose mixing can induce a Majorana mass as depicted in Fig. \ref{fig:nunu} (b). Their corresponding quantum numbers are listed in Table \ref{tab:coup}.

 \begin{table}[tbh]
\centering 
\begin{tabular}{|c|c|c|c|c|}\hline
Leptoquark & $(SU(3),SU(2))_{U(1)_Y}$ & $Q_{\text{EM}}$ & $B$ & $L$ \\ \hline
$S_{1/2}$ & $(3,2)_{1/6}$ & $(-1/3, 2/3)$ & $1/3$ & $-1$ \\ 
$S_{1}$ & $(3,3)_{-1/3}$ & $(2/3, -1/3, -4/3)$ & $1/3$ & $1$ \\ \hline
\end{tabular}
\caption{Quantum numbers of the leptoquarks with electric charge $2/3$ that are typically used to explain the $R_K$ anomaly.}
\label{tab:coup}
\end{table}

\vspace{-0.5cm}

To analyze effects on $R_K$ it is convenient to work with a flavor changing $|\Delta B| = |\Delta S| = 1$ effective Hamiltonian, integrating out the heavy degrees of freedom
\begin{align}
\mathcal{H}_{\text{eff}} = -4 \frac{G_F}{\sqrt{2}} V_{tb} V_{ts}^{\ast} \frac{\alpha_e}{4 \pi} \sum_i C_i \mathcal{O}_i\,, \label{heff}
\end{align}
where $G_F$ is the Fermi constant, $\alpha_e$ the electromagnetic coupling, $V_{ud}$ the Cabibbo-Kobayashi-Maskawa (CKM) matrix elements, and $C_i$the Wilson coefficients of their operators $\mathcal{O}_i$.

By adding $S_1$ and $S_{1/2}$ to the SM field content we can write down leptoquark couplings of the form 
\begin{align}
\mathcal{L}_{\text{LQ}} = \lambda^{R}_{S_{1/2}} \overline{d} P_L L S_{1/2} + \lambda^{L}_{S_1} \overline{Q^c} P_L i \tau_2 S_1^{\dagger} L\,,
\end{align}
which after Fierz rearrangement generate the effective (axial) vector operators
\begin{align} 
\mathcal{O}_9 &= \left[\overline{s} \gamma^{\mu} P_L b \right] \left[ \overline{l} \gamma_{\mu} l \right], \\ \mathcal{O}_{10} &=  \left[\overline{s} \gamma^{\mu} P_L b \right] \left[ \overline{l} \gamma_{\mu} \gamma_5 l \right],
\end{align}
and their chirality flipped counterparts $\mathcal{O}^{\prime}_{9,10}$ by interchanging the chiral projectors $P_L$ and $P_R$. A comparison with Eq. (\ref{heff}) yields $(l=e,\mu)$,
\begin{align} \begin{array}{lll}
S_{1}:\quad & C_{9}^{l} = - C_{10}^l &=  \frac{\pi}{\alpha_e} \frac{\left(\lambda^{L}_{sl}\right)^{\ast} \lambda^L_{bl}}{V_{tb} V_{ts}^{\ast}} \frac{\sqrt{2}}{2 M_{S_{1}}^2 G_F}\,, \\ 
S_{1/2}:\quad & C_{10}^{\prime l} = - C_{9}^{\prime l} &=  \frac{\pi}{\alpha_e}  \frac{\lambda^R_{sl} \left(\lambda^{R}_{bl}\right)^{\ast}}{V_{tb} V_{ts}^{\ast}} \frac{\sqrt{2}}{4 M_{S_{1/2}}^2  G_F}\,. 
\end{array}
\end{align}
As shown in \cite{Hirsch:1996qy,Kosnik:2012dj}, two leptoquarks sharing the same electric charge $Q$ will eventually mix through a coupling with the SM Higgs boson via
\begin{align}\begin{array}{ll}
V(S_i,H) =& -(M_{S_i}^2 - g_{S_i} H^{\dagger} H) S_i^{\dagger} S_i  \\ &+ h_S H i \tau_2 S_1 S_{1/2}^{\dagger} + \text{h.c.} \quad (i=1,1/2)\,.\end{array}
\end{align} 
The last term, in particular, accounts for the mixing and hence induces neutrino masses if $h_S~\neq~0$. On the other hand, if $h_S$ is large, the absolute neutrino mass scale generated by bottom- and leptoquark loops can also be too large, even to the extent of ruling out leptoquarks as an explanation for $R_K$.

The resulting leptoquark mass eigenstates are a mixture of flavor states with 
$Q_{\text{EM}}$ charges $2/3$, $-1/3$, and a distinct $-4/3$ state
\begin{align}\begin{array}{ll}
M^2_{2/3} &= \left( 
\begin{array}{cc}
M_{S_1}^2 - g_{S_1} v_{\text{SM}}^2  & h_S v_{\text{SM}}\\
h_S v_{\text{SM}} & M_{S_{1/2}}^2 - g_{S_{1/2}} v_{\text{SM}}^2
\end{array}
\right), \\   &= M^2_{-1/3}\,, \qquad
M^2_{-4/3} = M_{S_1}^2 - g_{S_1} v_{\text{SM}}^2\,.                                                                                                                                                                                                                                                      \end{array}                              \end{align}
The rotation angle $\alpha$ diagonalizing the $M_{2/3}^2$ matrix is determined by
\begin{align}
 \left( 
\begin{array}{c}
\tilde{S}_1 \\ 
\tilde{S}_2%
\end{array}%
\right) &= R \left( 
\begin{array}{c}
S_1 \\ 
S_{1/2}%
\end{array}%
\right)_{2/3}, ~R = \left( 
\begin{array}{cc}
\cos \alpha  & -\sin \alpha\\
\sin \alpha & \cos \alpha
\end{array}
\right) \label{massstates} \\
 &\text{and} \quad \tan 2\alpha = \frac{2 h_S v_{\text{SM}}}{M_{S_{1/2}}^2 - M_{S_{1}}^2}\,, \label{tanalpha}
\end{align}
where $\tilde{S_i}$ denotes the leptoquark mass eigenstates. The mixing shifts the absolute leptoquark masses and induces additional (pseudo-)scalar and tensor operators potentially affecting $\text{Br}(B_s \rightarrow ll)$ and $\text{Br}(B \rightarrow K ll)$. 
Their Wilson coefficients are
\begin{align}
C_S = C_P = \frac{\pi}{4 \alpha_e \sqrt{2} G_F V_{tb}V_{ts}^{\ast}} \cos \alpha \sin \alpha \nonumber \\
\times \lambda^R_{sl} \left(\lambda^{L}_{bl}\right)^{\ast} \left(\frac{1}{M_{\tilde{S}_2}^2} - \frac{1}{M_{\tilde{S}_1}^2}\right)
\end{align}
with similar expressions for $C^{\prime}_P$ and $C^{\prime}_S$ directly depending on the mixing angle $\alpha$. For $M_{S_{1/2}}^2 \gg M_{S_{1}}^2$ the coefficients simplify to
\begin{align}
C_S &= C_P \approx -\frac{\pi}{4 \alpha_e \sqrt{2} G_F V_{tb}V_{ts}^{\ast}} \frac{h_S v_{\text{SM}} \lambda^R_{sl} \left(\lambda^{L}_{bl}\right)^{\ast}}{M_{S_{1/2}}^2 M_{S_{1}}^2} \,, \label{CS} \\ 
-C^{\prime}_P &= C^{\prime}_S \approx-\frac{\pi}{4 \alpha_e \sqrt{2} G_F V_{tb}V_{ts}^{\ast}} \frac{h_S v_{\text{SM}} \lambda^L_{sl} \left(\lambda^{R}_{bl}\right)^{\ast}}{M_{S_{1/2}}^2 M_{S_{1}}^2}\,, \\
C_T &= (C_S + C^{\prime}_S)/4\,, \qquad C_{T_5} = (C_S - C^{\prime}_S)/4\,. \label{CT}
\end{align}
The Wilson coefficients $C_{S,P,T}$ are suppressed by a factor $h_S v_{\text{SM}}/|\Delta M_S^2|$, with $\Delta M_S^2 \equiv M_{S_{1/2}}^2 - M_{S_1}^2$, and therefore less relevant at low energies, but should be taken into account in the case of degenerate leptoquark masses. In the limit of small mixing, the shift of the leptoquark masses also becomes negligible for determining $R_K$. 

As shown, e.g., in \cite{Babu:2010vp}, the dimensional parameter $h_S$ cannot be arbitrarily large, but is in fact limited by the condition of positive leptoquark masses and the perturbativity of the theory to
\begin{align}
 h_S \leq M_{S_1} M_{S_{1/2}} / v_{\text{SM}}\,.
\end{align}

Consequently, taking into account only vector operators, the LHCb $R_K$ measurement implies ($1\,\sigma$) \cite{Hiller:2014yaa}
\begin{align}
&0.7 \lesssim \text{Re}\left[X^e-X^{\mu}\right] \lesssim 1.5\,, \quad \text{with} \label{rk} \\
&X^l = C_9^l + C_9^{\prime l} - (C_{10}^l + C_{10}^{\prime l}) \label{xl}\,.
\end{align}
Hence, with the combination of $S_1$ and $S_{1/2}$ we obtain
\begin{align} \begin{array}{ll} 
X^e - X^{\mu} &= \underbrace{\frac{\pi}{\sqrt{2} \alpha_e G_F V_{tb} V_{ts}^{\ast}}}_{\approx (25\,\text{TeV})^2}  \\ 
&\times \left[ \frac{2}{M_{S_1}^2} \left( \left(\lambda^{L}_{se}\right)^{\ast} \lambda^L_{be} - \left(\lambda^{L}_{s\mu}\right)^{\ast} \lambda^L_{b\mu} \right) \right. \\
 &- \left. \frac{1}{M_{S_{1/2}}^2} \left( \lambda^R_{se} \left(\lambda^{R}_{be}\right)^{\ast} - \lambda^R_{s\mu} \left(\lambda^{R}_{b\mu}\right)^{\ast} \right) \right]. 
\end{array} \label{xem}
\end{align}
By considering all constraints on leptoquark couplings to down-type quarks and charged leptons, the authors of \cite{Varzielas:2015iva} came up with a "data-driven" pattern that complies with all current bounds and can produce a visible signal in $b \rightarrow s ll$ processes:
\begin{align}
 \lambda_{ql} \sim \lambda_0 \left( 
\begin{array}{ccc}
\rho_d \kappa & \rho_d & \rho_d \\ 
\rho \kappa & \rho & \rho \\ 
\kappa & 1 & 1
\end{array}
\right) \label{pattern}
\end{align}
with the allowed parameter ranges (assuming $\lambda_0 \approx 1$)
\begin{align} \begin{array}{ll}
 \rho_d \lesssim 0.02\,,& \quad \kappa \lesssim 0.5, \quad 10^{-4} \lesssim \rho \lesssim 1\,, \\ &\frac{\kappa}{\rho} \lesssim 0.5\,, \quad  \frac{\rho_d}{\rho} \lesssim 1.6\,. \end{array}
\end{align}
The overall scale $\lambda_0$ is fixed by Eqs. (\ref{rk})-(\ref{xem}) and the leptoquark masses $M_{S_i}$, where $\lambda_0 \simeq \mathcal{O}(10^{-2})$ corresponds to light leptoquarks of a few TeV, while $\lambda_0 \approx 1$ implies heavy leptoquarks. 
Assuming for simplicity that both, $\lambda^L$ and $\lambda^R$ follow this pattern -- many other possibilities are plausible, too -- Eq. (\ref{xem}) simplifies to
\begin{align}\begin{array}{ll}
X^e - X^{\mu} &= \frac{\pi}{\sqrt{2} \alpha_e G_F V_{tb} V_{ts}^{\ast}} \\ &\times~ \lambda_0^2 \rho (\kappa^2 - 1) \left(\frac{2}{M_{S_{1}}^2} - \frac{1}{M_{S_{1/2}}^2}\right).\end{array}
\end{align}
This can be matched perfectly well to fit
Eq. (\ref{rk}) for a suitable choice of couplings and leptoquark masses. If, e.g., the leptoquark masses are strongly hierarchical with $M_{S_{1/2}} \gg M_{S_1}$, the mixing between them is minimized, thereby rendering additional (pseudo-)scalar and tensor operators negligible. $R_K$ is then almost exclusively determined by $S_{1}$. It is noteworthy that, if $S_{1/2}$ indeed contributes significantly to $R_K$, the measurement of $R_{K^{\ast}}$ will be a smoking gun of new physics since right-handed currents lead to deviations in the double ratio $\frac{R_{K^{\ast}}}{R_K} \neq 1$ \cite{Hiller:2014ula}.

\section{Generating neutrino masses \label{secmasses}}
If at least two leptoquarks with couplings to down-type quarks are present, a Majorana neutrino mass term as depicted in Fig. \ref{fig:nunu} (b) can be obtained at one-loop level. As pointed out in
 \cite{Buchmuller:1986zs,Hirsch:1996qy}, the leptoquark mixing with the Higgs boson will then generate a nonzero Majorana neutrino mass depending on the leptoquark couplings $\lambda^{L,R}$ and the mixing among the scalars. This is particularly interesting in models where a combination of two leptoquarks is considered to explain the measured $R_K$ ratio. 

Assuming that the scalar leptoquarks $S_1$ and $S_{1/2}$ are used simultaneously to reproduce $R_K$,
the contributions to the neutrino mass matrix read
\begin{align} 
 M^{\nu}_{i i^{\prime}} &= \frac{3}{16 \pi^2} \sum_{j=1,2} \sum_{k=d,s,b} m_k B_0(0,m_k^2,M_j^2) R_{j1} R_{j2} \nonumber \\
&\times \left[ \left(\lambda^R_{S_{1/2}}\right)_{k i} \left(\lambda^L_{S_{1}}\right)_{k i^\prime} + \left(\lambda^R_{S_{1/2}}\right)_{k i^\prime} \left(\lambda^L_{S_{1}}\right)_{k i} \right] \label{numass}
\end{align}
with $M_j$ being the mass of leptoquark $j$, and $R_{jl}$ the elements of the mixing matrix diagonalizing the leptoquark mass matrix. Following \cite{AristizabalSierra:2007nf}, we need consider only the finite part of the Passarino-Veltman function $B_0$
since the divergences cancel out due to $-R_{11}R_{12} = R_{21}R_{22}=\cos \alpha \sin \alpha$. In the relevant leptoquark mass range $1\,\text{TeV}\lesssim M_j \lesssim 50\,\text{TeV}$ $B_0$ is limited to 
\begin{align}
B_0(0,m_k^2,M_j^2) = \frac{m_k^2 \log(m_k^2) - M_j^2 \log (M_j^2)}{m_k^2 - M_j^2} \lesssim  8\,.
\end{align}

Starting from the data-driven pattern shown in Eq. (\ref{pattern}), we can estimate the absolute neutrino mass scale generated by the leptoquark couplings. The pattern is strongly hierarchical in terms of quark families since the light generations are tightly constrained by kaon phenomenology. Therefore, it is clear that the bottom quark will be the dominant contribution assuming that $\lambda^L$ and $\lambda^R$ have a similar structure. The latter does not have to be the case, and other plausible scenarios are discussed in the remainder of this paper. 

Considering only the dominant bottom quark loop we obtain
\begin{align}
M_{i i^{\prime}}^{\nu} &\approx \underbrace{\frac{3}{16 \pi^2} m_b \cos \alpha \sin \alpha \Delta B_0}_{\equiv a} \nonumber\\
&\times \left[ \left(\lambda^R_{S_{1/2}}\right)_{b i} \left(\lambda^L_{S_{1}}\right)_{b i^\prime} + \left(\lambda^R_{S_{1/2}}\right)_{b i^\prime} \left(\lambda^L_{S_{1}}\right)_{b i} \right], \\
\Delta B_0 &\equiv \left(B_0(0,m_b^2,M_{S_{1/2}}^2) - B_0(0,m_b^2,M_{S_{1}}^2)\right),
\end{align}
and after inserting Eq. (\ref{pattern})
\begin{align} 
M_{i i^{\prime}}^{\nu} \propto a \cdot \lambda_0^2 \left( 
\begin{array}{ccc}
\kappa^2 & \kappa & \kappa \\ 
\kappa & 1 & 1 \\ 
\kappa & 1 & 1
\end{array}
\right), \label{estimate}
\end{align}
while the neutrino mass eigenstates in terms of the leptoquark couplings are given by
\begin{align}
 m^{\nu}_1 &= 0, \\
 m^{\nu}_{2,3} &= \sum_{i} \lambda^L_{bi} \lambda^R_{bi} \pm \sqrt{\sum_{i} \left(\lambda^L_{bi}\right)^2 \sum_{i} \left(\lambda^R_{bi}\right)^2} \label{num}
\end{align}
with $\lambda^R_{S_{1/2}} \equiv \lambda^R$, $\lambda^L_{S_{1}} \equiv \lambda^L$ and $i=e,\mu,\tau$. In agreement with \cite{AristizabalSierra:2007nf}, one mass eigenvalue is exactly zero if either only down-type or up-type quarks generate neutrino masses.

Judging from Eqs. (\ref{estimate}) -- (\ref{num}), a small breaking of the $\mu - \tau$ symmetry of the original pattern is required to obtain two nonzero neutrino mass eigenvalues, and - if leptoquarks are the sole origin of lepton mixing - to reproduce the Pontecorvo-Maki-Nakagawa-Sakata (PMNS) mixing matrix observed in neutrino oscillations. 

Ignoring this problem for the moment, moreover
the rough estimate using the data-driven pattern yields an eigenvalue of order 1 times the bottom quark mass, which lies in the GeV range. Taking into account that $\lambda_0$ can be as small as $\mathcal{O}(10^{-2})$, we can suppress the neutrino mass scale by another four orders of magnitude. It then comes down to explaining the smallness of the parameter $a$ to reduce the mass scale by another five orders of magnitude. We can either attribute this to a tiny mixing of the leptoquarks with the SM Higgs, or a degeneracy among the leptoquark masses, for which the difference of Passarino-Veltman functions
\begin{align}
\Delta B_0 \approx \log\left[\frac{M_{S_{1/2}}^2}{M_{S_{1}}^2}\right] \qquad \left( m_b^2 \ll M_{S_i}^2 \right)
\end{align}
approaches zero. In this limit $\tan 2\alpha$ [cf. Eq. (\ref{tanalpha})] goes to infinity leading to maximal mixing among the leptoquarks. As a result $\cos \alpha \sin \alpha = 0.5$ will take its maximal but finite value. As stated earlier, however, with $|\Delta M_S^2|$ becoming small, additional scalar, pseudoscalar and tensor $b \rightarrow sll$ inducing operators become relevant that have to be taken into account in the calculation of $R_K$.

Interestingly, the mixing is minimized in the opposite scenario with strongly hierarchical leptoquark masses $M_{S_{1/2}}~\gg~M_{S_{1}}$. The extent is limited by the upper bound on the leptoquark mass $M_S \lesssim 50\,$TeV to comply with the phase measured in $B_s - \overline{B}_s$ mixing. $\Delta B_0 \approx 7.8$ then takes its maximal value, while $\tan 2\alpha$ is suppressed by a factor of up to 2500. Last but not least, one can also argue with the smallness of $h_S v_{\text{SM}}$ compared to $M_S^2$ to reduce the mixing by a few orders of magnitude.

In the following we concentrate on the alternative possibility to suppress the absolute neutrino mass by adopting different structures
for the coupling matrices $\lambda^L$ and $\lambda^R$.

\section{Alternative setups \label{secsetup}}
As stated above, $\lambda^L$ and $\lambda^R$ need not necessarily have the same pattern. Distinct couplings together with a slight breaking of the $\mu -\tau$ symmetry are actually favored by neutrino oscillation experiments requiring two nonzero neutrino mass eigenstates and large mixing angles. The latter dictate certain relations between the leptoquark couplings to explain the hierarchy of the mixing angles $\theta_{13} < \theta_{12} < \theta_{23}$, assuming that leptoquarks alone are responsible for leptonic mixing \cite{AristizabalSierra:2007nf}:
\begin{align}
\lambda^L_{qe} \lambda^R_{q\tau} + \lambda^L_{q\tau} \lambda^R_{qe} \ll \lambda^L_{q\mu} \lambda^R_{q\tau} + \lambda^L_{q\tau} \lambda^R_{q\mu}\,, \\
\lambda^L_{q\mu} \lambda^R_{q\tau} + \lambda^L_{q\tau} \lambda^R_{q\mu} \gtrsim \lambda^L_{q\mu} \lambda^R_{q\mu} - \lambda^L_{q\tau} \lambda^R_{q\tau}\,.
\end{align}
Furthermore, the couplings in $\lambda^L$ and $\lambda^R$ do not have to be the same order of magnitude. One or the other could have significantly smaller couplings to the SM fermions, effectively rendering the corresponding leptoquark redundant for the explanation of the $R_K$ anomaly, while at the same time reducing the absolute neutrino mass scale considerably.

In the case in which one does not demand that both leptoquarks contribute to $R_K$, a third possibility opens up that forces one of the leptoquarks to couple exclusively to lighter quark generations, e.g.,
\begin{align}
\lambda \simeq \left( 
\begin{array}{ccc}
\eta_{de} & \eta_{d\mu} & \eta_{d\tau} \\ 
0 & 0 & 0 \\ 
0 & 0 & 0
\end{array}
\right), \quad 
\lambda \simeq \left( 
\begin{array}{ccc}
0 & 0 & 0 \\ 
\eta_{se} & \eta_{s\mu} & \eta_{s\tau} \\ 
0 & 0 & 0
\end{array}
\right).
\end{align}
The $\eta_{sl}$ and $\eta_{dl}$ entries are constrained to be small by experimental data and will, therefore,
lower
 the neutrino mass scale. Moreover, the mass scale is further suppressed by the light quark masses $m_s, m_d \ll m_b$ now dominating the loop. Consequently, the requirements for the leptoquark-Higgs mixing are much more relaxed as opposed to the previous scenarios.

Finally, if one leptoquark couples solely to up-type quarks, neutrino masses can be generated only at two-loop level \cite{Babu:2010vp}. This mechanism sufficiently suppresses the neutrino mass scale, but is independent of the $R_K$ anomaly.

\vspace{0.5cm}
A quick numerical example will demonstrate that some reasonable choice of parameter values can indeed combine the $R_K$ measurement with the light neutrino mass scale at one-loop level. Starting from the patterns
\begin{align}
\lambda^L_{S_{1}} \sim \lambda_0 \left( 
\begin{array}{ccc}
\rho_d \kappa & \rho_d & \rho_d \\ 
\rho \kappa & \rho & \rho \\ 
\kappa & 1 & 1
\end{array}
\right), \quad \lambda^R_{S_{1/2}} \sim \left( 
\begin{array}{ccc}
\eta_{de} & \eta_{d\mu} & \eta_{d\tau} \\ 
\eta_{se} & \eta_{s\mu} & \eta_{s\tau} \\ 
\eta_{be} & \eta_{b\mu} & \eta_{b\tau}
\end{array}
\right),
\end{align}
the neutrino mass matrix $M^{\nu}_{i i^{\prime}}$ is approximately given by
\begin{align}
M^{\nu}_{i i^{\prime}} \approx a &\left( 
\begin{array}{ccc}
2 \eta_{be} \kappa & \kappa \eta_{b\mu} + \eta_{be} & \kappa \eta_{b\tau} + \eta_{be} \\ 
\kappa \eta_{b\mu} + \eta_{be} & 2 \eta_{b\mu} & \eta_{b\tau} + \eta_{b\mu} \\ 
\kappa \eta_{b\tau} + \eta_{be} & \eta_{b\tau} + \eta_{b\mu} & 2 \eta_{b\tau}
\end{array}
\right)\label{general} \\ \text{with} \quad
a&=\frac{3 \lambda_0}{16 \pi^2} m_b \cos \alpha \sin \alpha \log\left[\frac{M_{S_{1/2}}^2}{M_{S_{1}}^2}\right].
\end{align}
In the limit $M_{S_1} \approx 1\,$TeV $\ll M_{S_{1/2}} \approx 50\,$TeV, $S_1$ will be the dominant contribution to $R_K$. Choosing $\lambda_0, \kappa$ and $\rho \sim \mathcal{O}(\epsilon)$ with $\epsilon \approx 0.2$ will then comply with flavor physics precision measurements and yield $0.7 \lesssim \text{Re}\left[X^e-X^{\mu}\right] \lesssim 1.5$ implied by $R_K$ at $1\,\sigma$.

By using a specific ansatz for the leptoquark couplings, e.g., $
\eta_{ql} \simeq m_q m_l/v_{\text{SM}}^2$,
we are able to pin down the neutrino masses. Such hierarchical patterns are motivated by and easily obtained in Froggatt-Nielsen-type flavor models, where the fermion mass hierarchies are explained through an additional $U(1)_{\text{FN}}$ family symmetry \cite{Froggatt:1978nt,AristizabalSierra:2007nf,Chankowski:2005qp}. 

As an example, let us consider the fermion and leptoquark charges listed in Table \ref{tab:FN}, which can reproduce the SM fermion mass hierarchies ($\epsilon \approx 0.2$)
\begin{align}
 \label{e:rel}
 \begin{array}{lllllllllll}
  m_u & : & m_{c} & : & m_{t} & \approx & \epsilon^8 & : & \epsilon^4 & : & 1~,\\
  m_d & : & m_{s} & : & m_{b} & \approx & \epsilon^7 & : & \epsilon^5 & : & \epsilon^3~,\\
  m_e & : & m_{\mu} & : & m_{\tau} & \approx & \epsilon^9 & : & \epsilon^5 & : & \epsilon^3~,\\
 \end{array}
\end{align}
as well as the hierarchy of the CKM matrix up to $\mathcal{O}(1)$ coefficients
\begin{align}
V_{us} \simeq \epsilon\,, \quad V_{ub} \simeq \epsilon^3\,, \quad V_{cb} \simeq \epsilon^2\,.
\end{align}

 \begin{table}[tbh]
\centering 
\begin{tabular}{|c||c|c|c|c|c|c|c|c|c|}\hline
Field & $\overline{Q}_1$ & $\overline{Q}_2$ & $\overline{Q}_3$ & $d$ & $s$ & $b$ & $u$ & $c$ & $t$ \\ \hline
$Q(U(1)_{\text{FN}})$ & -2 & -1 & 0 & 9 & 6 & 3 & 10 & 5 & 0 \\ \hline
\end{tabular}

\vspace{0.2cm}
\begin{tabular}{|c||c|c|c|}\hline
Field & $\overline{L}_1$ & $\overline{L}_2$ & $\overline{L}_3$ \\ \hline
$Q(U(1)_{\text{FN}})$ & $-Q(L_3)-1$ & $-Q(L_3)$ & $-Q(L_3)$ \\ \hline
Field & $e$ & $\mu$ & $\tau$ \\ \hline
$Q(U(1)_{\text{FN}})$ & $Q(L_3) + 10$ & $Q(L_3) + 5$ & $Q(L_3) + 3$ \\ 
\hline
\end{tabular}

\vspace{0.2cm}
\begin{tabular}{|c||c|c|}\hline
Field & $S_1$ & $S_{1/2}$ \\ \hline
$Q(U(1)_{\text{FN}})$ & $Q(L_3)-1$ & $11-Q(L_3)$ \\ \hline
\end{tabular}
\caption{Possible $U(1)$ quantum numbers consistent with the SM fermion mass hierarchies and $V_{\text{CKM}}$ to obtain the patterns discussed in Eq. (\ref{FNpatt}).}
\label{tab:FN}
\end{table}

The resulting leptoquark patterns are the following:
\begin{align} 
\lambda^L_{S_1} \simeq \left( 
\begin{array}{ccc}
\epsilon^4 & \epsilon^3 & \epsilon^3 \\ 
\epsilon^3 & \epsilon^2 & \epsilon^2 \\ 
\epsilon^2 & \epsilon & \epsilon
\end{array}
\right),\quad \lambda^R_{S_{1/2}} \simeq \left( 
\begin{array}{ccc}
\epsilon^3 & \epsilon^2 & \epsilon^2 \\ 
\epsilon^6 & \epsilon^5 & \epsilon^5 \\ 
\epsilon^9 & \epsilon^8 & \epsilon^8
\end{array}
\right), \label{FNpatt}
\end{align}
which correspond to $\lambda_0 \approx \epsilon$, $\rho \approx \epsilon, \rho_d \approx \epsilon^2$ and $\kappa \approx \epsilon$. Note that since $S_1$ couples to $Q_i$, but $S_{1/2}$ couples to $\overline{d}$, one of the leptoquark patterns is inverse hierarchical in terms of quark generations, i.e., coupling strongly to $d$ and weakly to $b$ quarks. Hence, if we accommodate two leptoquarks simultaneously in a FN flavor symmetry, one of the leptoquarks always suppresses the strong couplings of the other, consequently leading to naturally light neutrino masses. Since
\begin{align} 
 &m_q \lambda^R_{q i} \lambda^L_{q i^\prime} \approx m_b \epsilon^{9} \qquad (q=b,s,d)\\
 &\text{with} \quad m_s \approx m_b \epsilon^2 \quad \text{and} \quad m_d \approx m_b \epsilon^4\,, \nonumber
\end{align}
all quark loops contribute equally to Eq. (\ref{numass}), resulting in an additional factor of $3$ in our estimate of the neutrino mass scale.
Taking $h_S = 1\,$TeV, we find
\begin{align}
m^{\nu}_3 \approx 0.30\,\text{eV}, \qquad m^{\nu}_2 \approx 0.01\,\text{eV}, \qquad m^{\nu}_1 = 0,
\end{align}
which approximates the expected neutrino mass scale 
quite well. The structure shown in Eq. (\ref{general}) also guarantees three nonzero mixing angles allowing one to declare that leptoquarks are the sole origin of neutrino masses and mixings. Obtaining the observed mass squared differences $\Delta m_{\text{atm}}^2$ and $\Delta m_{\text{sol}}^2$ and the exact mixing angle values, however, would require a precise fit and an explicit model realization, which is discussed here.

With such inverse hierarchical patterns as in Eq. (\ref{FNpatt}) one must keep in mind constraints from low energy experiments, where typically the strongest bound comes from rare kaon decay data \cite{Davidson:1993qk}
\begin{align}
 |\lambda_{d\mu} \lambda_{s\mu}^{\ast}| \lesssim \frac{M_S^2}{(183\,\text{TeV})^2}\,.
\end{align}
Further constraints on $\Delta L=2$ 
lepton number violating
 leptoquark couplings also arise from neutrinoless double beta ($0\nu\beta\beta$) experiments, which can be even more stringent than LHC searches \cite{Hirsch:1996ye,Helo:2013ika}. This is particularly interesting since one of the leptoquarks interacts strongly with the first quark generation. The $0\nu\beta\beta$ decay requires leptoquark couplings to up-type quarks that in our framework are only provided by $S_1$. Consequent mixing with $S_{1/2}$ then generates the essential operator \cite{Hirsch:1996ye}
\begin{align} 
 C \left[\overline{\nu} P_R e^c \right] \left[\overline{u} P_R d \right]
 \end{align}
 with
 \begin{align}
C = \lambda^L_{S_1} \lambda^R_{S_{1/2}}
 \left( \frac{ R_{11} R_{12}}{M_{S_1}^2} + 
\frac{ R_{21} R_{22} }{M_{S_{1/2}}^2} \right)\,,
\end{align}
which is due to our limited leptoquark content the only contribution to the $0\nu\beta\beta$ decay. Hence, the expression for the half-life simplifies to
\begin{align}
 T_{1/2}^{0\nu\beta\beta} = \left(|\mathcal{M}_{GT}|^2 \tilde{a} \frac{2}{G_F^2} C^2 \right)^{-1},
 \end{align}
where $\mathcal{M}_{GT}$ is the nuclear matrix element and $\tilde{a}$ a function of the electron mass and the nuclear radius. For$~^{76}$Ge, $|\mathcal{M}_{GT}|^2  \tilde{a} = 6.52 \times 10^{-10}$ was obtained numerically in \cite{Hirsch:1996ye} and afterwards corrected by a factor of 4 in \cite{Pas:1998nn}. 

The best bound on the half-life of $~^{76}$Ge is currently provided by GERDA \cite{Agostini:2013mzu} with
\begin{align}
 T_{1/2}^{0\nu\beta\beta}(^{76}\text{Ge}) > 2.1 \times 10^{25}\,\text{yr}.
\end{align}
Using the leptoquark patterns shown in Eq. (\ref{FNpatt}), we obtain $\lambda^L_{11} \lambda^R_{11} = \epsilon^{7}$ and consequently
\begin{align}
 T_{1/2}^{0\nu\beta\beta}(^{76}\text{Ge}) \approx \frac{2.5 \times 10^{21}\,\text{yr}}{(\cos \alpha \sin \alpha)^2} \frac{M_{S_{1/2}}^4 M_{S_1}^4}{\Delta M_S^4 (\text{TeV})^4}\,.
\end{align}
For our numerical example with $M_{S_{1/2}} \simeq 50\,$TeV, $M_{S_1}\simeq 1\,$TeV and $h_S = 1\,$TeV, the $0\nu\beta\beta$ half-life results in
\begin{align}
T_{1/2}^{0\nu\beta\beta}(^{76}\text{Ge}) \approx  9.7 \times 10^{29}\,\text{yr}\,.
\end{align}

If the hierarchy of $\lambda^R_{S_{1/2}}$ is even stronger, i.e., $m_d \approx m_b \epsilon^5$, it is possible to suppress the neutrino mass scale while getting dangerously close to the current $0\nu\beta\beta$ bound. In the scenario of almost degenerate leptoquark masses, $M_{S_{1}} \simeq 1\,$TeV and $M_{S_{1/2}} \simeq 2\,$TeV, $S_{1/2}$ still does not contribute to $R_K$ due to its tiny $s$ and $b$ quark couplings. Yet, as opposed to the previous example, the mixing is now significantly enhanced with $\alpha \approx 0.01$ for $h_S = 0.1 \,$TeV. The extra suppression of $\lambda^R_{S_{1/2}}$ then compensates for the strong leptoquark mixing to keep the neutrino masses in the eV range. These values thus yield a half-life of
\begin{align}
T_{1/2}^{0\nu\beta\beta}(^{76}\text{Ge}) \approx  2.7 \times 10^{26}\,\text{yr}\,,
\end{align}
which lies within the expected sensitivity of GERDA phase II.
We conclude that the $0 \nu \beta\beta$ limit is respected thanks to the small leptoquark mixing, but could in principle allow an observation of $0\nu\beta\beta$ decay in the 
near future. 

Another interesting and important task would be to study the washout effect on the
baryon asymmetry of the Universe but this is beyond the scope of the present analysis \cite{Deppisch:2015yqa}.

\section{Conclusion \label{secconc}}
Two scalar leptoquarks can generate neutrino masses at one-loop level and simultaneously explain the $2.6\,\sigma$ anomaly in $R_K$ recently announced by LHCb. This is possible if these leptoquarks mix weakly with the SM Higgs boson to induce a $\Delta L=2$ effective Majorana mass term.

We have exemplified this using the $(3,2)_{1/6}$ and $(3,3)_{-1/3}$ leptoquark representations.

Small Higgs-leptoquark mixing will lead to naturally small neutrino masses and protect the tightly constrained 
$B$ decays from extra flavor changing $|\Delta B|=|\Delta S|=1$
scalar and tensor operators. At the same time, leptoquark couplings to $b$ and $s$ quarks may still be strong enough to produce visible signals in $B \rightarrow K ll$ decays, with leptoquark masses confined to the testable region $1\,\text{TeV} \lesssim M \lesssim 50\,\text{TeV}$. Suitable patterns can be obtained by embedding the leptoquarks together with the SM fields in Froggatt-Nielsen-type flavor models.


\begin{thebibliography}{9}
\bibitem{Hiller:2003js} 
  G.~Hiller and F.~Kruger,
  Phys.\ Rev.\ D {\bf 69}, 074020 (2004)
  doi:10.1103/PhysRevD.69.074020
  [hep-ph/0310219].



\bibitem{Aaij:2014ora} 
  R.~Aaij {\it et al.} [LHCb Collaboration],
  Phys.\ Rev.\ Lett.\  {\bf 113}, 151601 (2014)
  doi:10.1103/PhysRevLett.113.151601
  [arXiv:1406.6482 [hep-ex]].



\bibitem{Hiller:2014yaa} 
  G.~Hiller and M.~Schmaltz,
  Phys.\ Rev.\ D {\bf 90}, 054014 (2014)
  doi:10.1103/PhysRevD.90.054014
  [arXiv:1408.1627 [hep-ph]].



\bibitem{Hiller:2014ula} 
  G.~Hiller and M.~Schmaltz,
  JHEP {\bf 1502}, 055 (2015)
  doi:10.1007/JHEP02(2015)055
  [arXiv:1411.4773 [hep-ph]].



\bibitem{Varzielas:2015iva} 
  I.~de Medeiros Varzielas and G.~Hiller,
  JHEP {\bf 1506}, 072 (2015)
  doi:10.1007/JHEP06(2015)072
  [arXiv:1503.01084 [hep-ph]].



\bibitem{Sahoo:2015qha} 
  S.~Sahoo and R.~Mohanta,
  arXiv:1507.02070 [hep-ph].



\bibitem{Sahoo:2015fla} 
  S.~Sahoo and R.~Mohanta,
  arXiv:1509.06248 [hep-ph].



\bibitem{Alonso:2015sja} 
  R.~Alonso, B.~Grinstein and J.~M.~Camalich,
  JHEP {\bf 1510}, 184 (2015)
  doi:10.1007/JHEP10(2015)184
  [arXiv:1505.05164 [hep-ph]].



\bibitem{deBoer:2015boa} 
  S.~de Boer and G.~Hiller,
  arXiv:1510.00311 [hep-ph].



\bibitem{Becirevic:2015asa} 
  D.~Bečirević, S.~Fajfer and N.~Košnik,
  Phys.\ Rev.\ D {\bf 92}, no. 1, 014016 (2015)
  doi:10.1103/PhysRevD.92.014016
  [arXiv:1503.09024 [hep-ph]].



\bibitem{Calibbi:2015kma} 
  L.~Calibbi, A.~Crivellin and T.~Ota,
  Phys.\ Rev.\ Lett.\  {\bf 115}, 181801 (2015)
  doi:10.1103/PhysRevLett.115.181801
  [arXiv:1506.02661 [hep-ph]].



\bibitem{Gripaios:2014tna} 
  B.~Gripaios, M.~Nardecchia and S.~A.~Renner,
  JHEP {\bf 1505}, 006 (2015)
  doi:10.1007/JHEP05(2015)006
  [arXiv:1412.1791 [hep-ph]].



\bibitem{Hirsch:1996ye} 
  M.~Hirsch, H.~V.~Klapdor-Kleingrothaus and S.~G.~Kovalenko,
  Phys.\ Rev.\ D {\bf 54}, 4207 (1996)
  doi:10.1103/PhysRevD.54.R4207
  [hep-ph/9603213].



\bibitem{Mahanta:1999xd} 
  U.~Mahanta,
  Phys.\ Rev.\ D {\bf 62}, 073009 (2000)
  doi:10.1103/PhysRevD.62.073009
  [hep-ph/9909518].



\bibitem{AristizabalSierra:2007nf} 
  D.~Aristizabal Sierra, M.~Hirsch and S.~G.~Kovalenko,
  Phys.\ Rev.\ D {\bf 77}, 055011 (2008)
  doi:10.1103/PhysRevD.77.055011
  [arXiv:0710.5699 [hep-ph]].



\bibitem{Babu:2010vp} 
  K.~S.~Babu and J.~Julio,
  Nucl.\ Phys.\ B {\bf 841}, 130 (2010)
  doi:10.1016/j.nuclphysb.2010.07.022
  [arXiv:1006.1092 [hep-ph]].



\bibitem{Kohda:2012sr} 
  M.~Kohda, H.~Sugiyama and K.~Tsumura,
  Phys.\ Lett.\ B {\bf 718}, 1436 (2013)
  doi:10.1016/j.physletb.2012.12.048
  [arXiv:1210.5622 [hep-ph]].



\bibitem{Cai:2014kra} 
  Y.~Cai, J.~D.~Clarke, M.~A.~Schmidt and R.~R.~Volkas,
  JHEP {\bf 1502}, 161 (2015)
  doi:10.1007/JHEP02(2015)161
  [arXiv:1410.0689 [hep-ph]].



\bibitem{Sierra:2014rxa} 
  D.~Aristizabal Sierra, A.~Degee, L.~Dorame and M.~Hirsch,
  JHEP {\bf 1503}, 040 (2015)
  doi:10.1007/JHEP03(2015)040
  [arXiv:1411.7038 [hep-ph]].



\bibitem{Helo:2015fba} 
  J.~C.~Helo, M.~Hirsch, T.~Ota and F.~A.~P.~d.~Santos,
  JHEP {\bf 1505}, 092 (2015)
  doi:10.1007/JHEP05(2015)092
  [arXiv:1502.05188 [hep-ph]].



\bibitem{Davidson:1993qk} 
  S.~Davidson, D.~C.~Bailey and B.~A.~Campbell,
  Z.\ Phys.\ C {\bf 61}, 613 (1994)
  doi:10.1007/BF01552629
  [hep-ph/9309310].



\bibitem{Hirsch:1996qy} 
  M.~Hirsch, H.~V.~Klapdor-Kleingrothaus and S.~G.~Kovalenko,
  Phys.\ Lett.\ B {\bf 378}, 17 (1996)
  doi:10.1016/0370-2693(96)00419-4
  [hep-ph/9602305].



\bibitem{Kosnik:2012dj} 
  N.~Kosnik,
  Phys.\ Rev.\ D {\bf 86}, 055004 (2012)
  doi:10.1103/PhysRevD.86.055004
  [arXiv:1206.2970 [hep-ph]].



\bibitem{Buchmuller:1986zs} 
  W.~Buchmuller, R.~Ruckl and D.~Wyler,
  Phys.\ Lett.\ B {\bf 191}, 442 (1987)
  [Phys.\ Lett.\ B {\bf 448}, 320 (1999)].
  doi:10.1016/0370-2693(87)90637-X



\bibitem{Froggatt:1978nt} 
  C.~D.~Froggatt and H.~B.~Nielsen,
  Nucl.\ Phys.\ B {\bf 147}, 277 (1979).
  doi:10.1016/0550-3213(79)90316-X



\bibitem{Chankowski:2005qp} 
  P.~H.~Chankowski, K.~Kowalska, S.~Lavignac and S.~Pokorski,
  Phys.\ Rev.\ D {\bf 71}, 055004 (2005)
  doi:10.1103/PhysRevD.71.055004
  [hep-ph/0501071].



\bibitem{Helo:2013ika} 
  J.~C.~Helo, M.~Hirsch, H.~Päs and S.~G.~Kovalenko,
  Phys.\ Rev.\ D {\bf 88}, 073011 (2013)
  doi:10.1103/PhysRevD.88.073011
  [arXiv:1307.4849 [hep-ph]].



\bibitem{Pas:1998nn} 
  H.~Päs, M.~Hirsch and H.~V.~Klapdor-Kleingrothaus,
  Phys.\ Lett.\ B {\bf 459}, 450 (1999)
  doi:10.1016/S0370-2693(99)00711-X
  [hep-ph/9810382].



\bibitem{Agostini:2013mzu} 
  M.~Agostini {\it et al.} [GERDA Collaboration],
  Phys.\ Rev.\ Lett.\  {\bf 111}, no. 12, 122503 (2013)
  doi:10.1103/PhysRevLett.111.122503
  [arXiv:1307.4720 [nucl-ex]].



\bibitem{Deppisch:2015yqa} 
  F.~F.~Deppisch, J.~Harz, M.~Hirsch, W.~C.~Huang and H.~Päs,
  Phys.\ Rev.\ D {\bf 92}, no. 3, 036005 (2015)
  doi:10.1103/PhysRevD.92.036005
  [arXiv:1503.04825 [hep-ph]].




\end{thebibliography}
\end{document}